# Collective Pinning and Vortex Dynamics in type II superconducting thin films with Varying Magnetic Field at T → 0

Yu Wu[2,3†], Liangliang Guo[1†], Renfei Wang[1], Jiawei Guo[1], Shuang Jia[1,6], Mingliang Tian[4,5], Xiaobo Lu[1], Hangwen Guo[2,6*], Jian Shen[2,3,6*], Yang Liu[1,6*]

[1]International Center for Quantum Materials, Peking University, Beijing, 100871, China

[2]State Key Laboratory of Surface Physics and Institute for Nanoelectronic Devices and Quantum Computing, Fudan University, Shanghai 200433, China

[3]Department of Physics, Fudan University, Shanghai, China, 200433

[4]Anhui Province Key Laboratory of Low-Energy Quantum Materials and Devices, High Magnetic Field Laboratory, HFIPS, Chinese Academy of Sciences, Hefei, Anhui 230031, China

[5]School of Physics and Optoelectronics Engineering, Anhui University, Hefei 230601, China

[6]Hefei National Laboratory, Hefei 230088, China

*Corresponding authors: liuyang02@pku.edu.cn (Y. L.); shenj5494@fudan.edu.cn (S. J.); hangwenguo@fudan.edu.cn (H. G.)

† These authors contributed equally to this work.

**Abstract:** A perpendicular magnetic field penetrating a thin type-II superconductor slab produces vortices, with one vortex per flux quantum, h/2e. The vortices interact repulsively and form an ordered array (Abrikosov lattice) in clean systems, while strong disorder changes the lattice into a vortex glass. Here we investigate type-II superconducting films (PdBi2 and NbSe2) with surface acoustic waves (SAWs) at mK temperature. When sweeping the magnetic field at an extremely slow rate, we observe a series of spikes in the attenuation and velocity of the SAW, on average separated in field by approximately $H_{c1}$. We suspect the following scenario: The vortex-free region at the edges of the film produces an edge barrier across which the vortices can enter or leave. When the applied field changes, the induced supercurrents flowing along this edge region lowers this barrier until there is an instability. At that point, vortices avalanche into (or out of) the bulk and change the vortex crystal, suggested by the sharp jump in each such spike. The vortices then gradually relax to a new stable pinned configuration, leading to a ~30s relaxation after the jump. Our observation enriches the limited experimental evidence on the important topic of real-time vortex dynamics in superconductors.

**Main Text:**

The vortices and their dynamics are one of the most important research topics in type II superconductors. A super-current ring traps one magnetic flux quanta at its center, and expands the critical field to $H_{c2}$[1–3]. They will be pinned by disorder[1,4,5], and these charge neutral objects experiences Lorentz force from current flowing through the sample, leading to finite dissipation below the expected critical current[6,7]. Their motion can lead to other phenomena such as radio frequency (RF) absorption[8–10] and decoherence of superconducting qubits[11,12], etc. Nevertheless, the vortices strongly interact with each other and tend to form an ordered array (Abrikosov lattice) whose collective pinning mechanism suppresses their motion[1,13,14]. The collective vortex dynamics is highly fragile to external perturbations[15-17] so that, despite its importance, experimental observation is limited. The existing works report either at low magnetic fields when vortices distribute sparsely[18-20], or at critical conditions where the motion of vortex causes significant change in superconductivity[8–10]. In thin layer cases, these tube-like vortex objects can be viewed as strongly interacting two dimensional pancake-like quasiparticles that form crystal, glass or liquid phases depending on the strength of disorder and temperature[1,21,22].

Surface acoustic wave (SAW) is an acousto-mode confined at the solid surface. It can be excited and detected by interdigital transducers (IDTs) fabricated on piezoelectric substrates[23]. It can interact with proximate material through either strain or piezoelectric field, resulting in its attenuation or velocity shift[24,25]. Early SAW studies on superconductors mainly focus near the critical temperatures, when the large number of thermally excited quasiparticles can generate a detectable signal if stimulated by SAW[26,27]. However, the large SAW excitation power and thermal fluctuation agitate the vortex and thus cloud the intrinsic vortex dynamics.

In this work, we investigate the thin type-II superconducting films (PdBi$_2$ and NbSe$_2$) with nW-level SAW at mK temperature when the excited quasiparticles vanish exponentially[1,28]. Remarkably, we observe an anomalously large SAW energy loss $\Gamma$ along with an enhanced

SAW velocity $\bar{\eta} = \Delta v/v$, quite likely caused by the energy dissipation due to the current-driven vortex motion. In particular, at the steady-field limit ($dB/dt \to 0$), $\eta$ becomes nearly flat except a plateau feature below the critical magnetic field ($H_{c1}$) which possibly signals zero vortex density (Meissner effect). Strikingly, $\Gamma$ and $\bar{\eta}$ exhibit oscillations during field sweeps which turn to event like spikes at extremely small $dB/dt$. Each spike consists a first-order jump followed by a slow decay, possibly related to a collapse and re-crystallization of the vortex lattice. The spikes persist at high field up to $H_{c2}$, but only at low temperatures. This phenomenon is extremely fragile because factors such as faster sweep rate, a flowing current, or raising temperature will diminish the oscillations while the superconductivity remains unchanged.

Our sample is a ~ 90 nm single crystal $\beta$-PdBi$_2$ flake, and the SAW resonance frequency is 839.6 MHz (see Fig.1(a) and (b)). Fig. 1(c) shows the magnetoresistance (the light blue curve) and the measured relative SAW average velocity shift $\bar{\eta} = \frac{v}{\Delta v}$ in reference to its value at high magnetic field when the sample is metallic (non-superconducting). The velocity decreases upon interacting with quasiparticles at the well-defined Fermi surface. When the magnetic field is smaller than $H_{c2}$, the sample becomes superconducting. The quasiparticle gap at the Fermi surface suppresses this interaction, so that $\bar{\eta}$ increases[29,30]. We define the SAW absorption as $\Gamma = -20 log\left(\frac{A(B)}{A_0}\right)$, where $A(B)$ is the measured SAW amplitude and $A_0$ is its value at high magnetic field. According to the relaxation model[31, 32] and BCS theory[33], $\Gamma$ should be negligible either when the sample is metallic with a large conductivity, or when the coupling between SAW and quasiparticles are suppressed by the superconductor gap near zero field. The vortices at $B > B_{c1}$ slow down the SAW, which should be consistent with the experimentally observed enhancement of $\bar{\eta}$ at $B < B_{c1} \cong 6$ mT when the vortex density is zero, as shown in Fig. 2(d)[34].

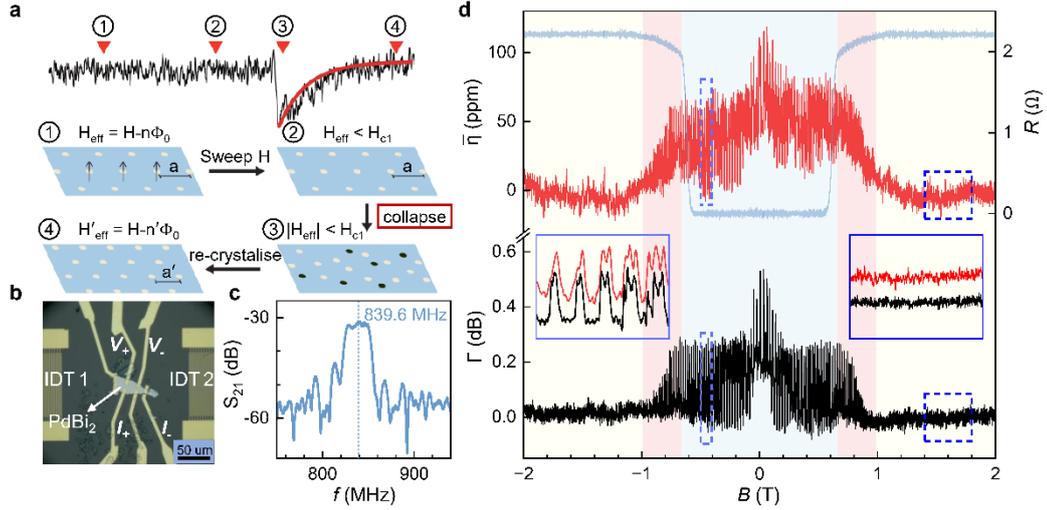

**Figure 1 | Principle of the measurement, velocity shift and attenuation. a,** Schematic diagram of a single event. At extremely low temperatures, vortices form a stable lattice structure. As the external magnetic field varies, the lattice remains unchanged due to collective pinning while the effective magnetic field experienced by the superconducting phase outside the vortices core $H_{eff}$ varies. When $H_{eff}$ reach the lower critical field $H_{c1}$, the Meissner effect fails, forcing new vortices to emerge and the vortex lattice to collapse. Subsequently, it slowly relaxes to form a new lattice with different lattice constant. **b,** Photo of the $PdBi_2$ device. $PdBi_2$ thin film is transferred to the middle of IDT1 and IDT2, where IDT2 is used to excite SAW and IDT1 is used to receive SAW. Transport electrodes are evaporated on top of $PdBi_2$ and the used four contacts have been labeled. **c,** $S_{21}$ parameter of $PdBi_2$ device. The center frequency is 839.6 MHz. **d,** SAW velocity shift $\bar{\eta}$ and attenuation $\Gamma$ vs. magnetic field. Both figures show characteristic SAW signal when $PdBi_2$ is in superconducting state, which is defined by $PdBi_2$'s magnetoresistance (light-blue curve). Insets show the curve details of two field ranges (light-blue and blue dashed boxes), where oscillation will appear in low field and disappear in high field.

The most striking feature in Fig. 1(c) and (d) is the oscillating behavior of $\bar{\eta}$ and $\Gamma$. These pronounced signals are not observed in the blank control device[35], confirming that the signal originates from the superconductor sample. The appearance of the oscillation coincides with the superconductivity, i.e., it emerges at the onset of superconductivity when the sample resistivity starts to decrease (the light red shade). Its amplitude saturates when the sample resistivity vanishes (the light blue region). Inside the oscillating region, a general increase of $\Gamma$ is seen when $\bar{\eta}$ increases. The generation of the supercurrent $J_s$, governed by the London equation:

$\frac{dJ_S}{dt} = \frac{1}{\mu_0 \lambda^2} E$, does not dissipate energy from the SAW by itself, but enhances the SAW velocity by its inductive response. On the other hand, the stimulated $J_S$ poses a Lorentz force ($F_L = J_S \times \Phi_0 k$) to the vortices. Once depinned, the vortex motion dissipates energy and cause a finite energy-loss of SAW[36–39]. Therefore, the piezo-electric field induced J oscillation can enhance the SAW velocity and dissipate its energy. This has been reported by previous studies, where SAW does not interact with cooper-pairs, but can be attenuated by the vortex lattice[26,27,40,41].

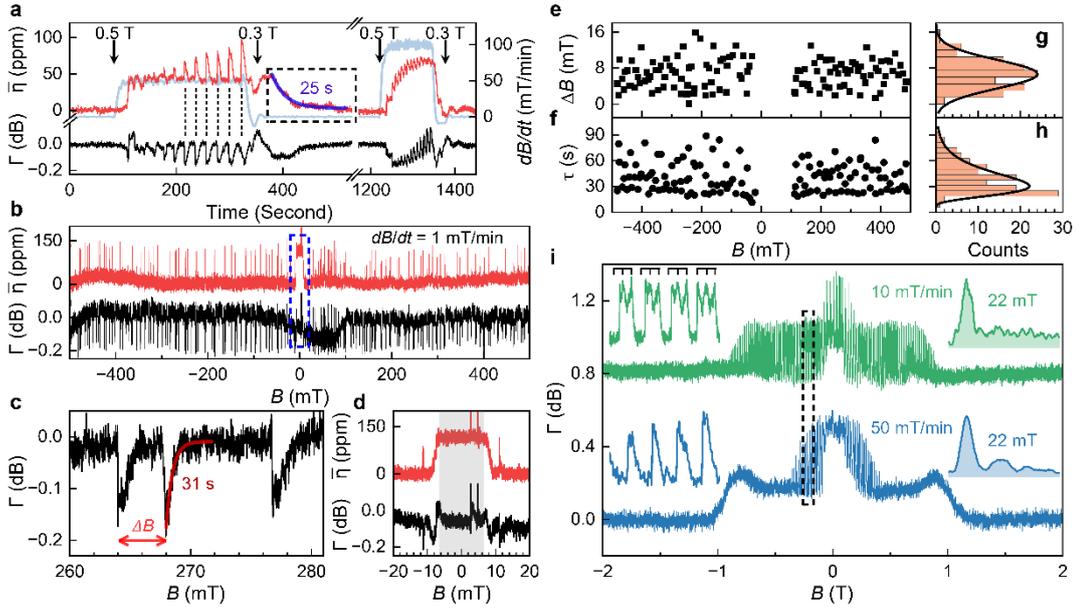

**Figure 2 | Magnetic field-dependent oscillations and quasi-static processes. a,** $\bar{\eta}$ and $\Gamma$ vs. time when B is stable and swept from 0.5 T to 0.3 T at 50 and 100 mT/min sweep rate (light-blue curve). **b,** $\bar{\eta}$ and $\Gamma$ when $B$ sweeps from -500 to 500 mT at 1 mT/min rate, the nearly periodic oscillation becomes discrete spikes, whose detailed behavior is plotted as the black curve in **c**. The spike-interval is $\Delta B$ and each spike's relaxation time $\tau$ is deduced from exponential decay fitting. **d,** $\bar{\eta}$ shows a plateau below $H_{c1}$. (e-h) Both $\Delta B$ and $\tau$ show no correlation with $B$ and their statistics can be fitted with normal and log-normal distribution. (i) $\Gamma$ vs. $B$ at different sweeping rates. Left insets show the expanded oscillation of data in the black box. When rate increases from 10 to 50 mT/min, the spike behavior will be weakened and gradually merge in group of three, forming a B-periodic oscillation whose frequency is ∼ 22 mT, shown in right insets' FFT diagrams. The overall attenuation curve will deform due to the merging of spikes.

These oscillations can only be seen upon sweeping magnetic field, as shown in Fig. 2(a). $\bar{\eta}$ and $\Gamma$ remains zero in reference to their value at high magnetic field above $B_{c2}$ if the magnetic field persists at either 0.3 or 0.5T. As soon as we sweep the magnetic field at 50 mT/min, $\bar{\eta}$ increases steadily by about 50 ppm while $\Gamma$ remains roughly unchanged. Meanwhile, oscillations appear in $\bar{\eta}$ and corresponding minimums are seen in $\Gamma$. When the field sweeping stops, $\bar{\eta}$ decreases to zero exponentially with a time constant of about 25s (see the blue curve inside the dashed box). If we double the sweeping speed, the amplitudes of both $\bar{\eta}$ and $\Gamma$ oscillations become smaller and $\Gamma$ first becomes negative and then increases to be positive. Because neither the normal metal nor the superconductor is expected to have absorption, the negative $\Gamma$ actually suggests an amplification of the SAW by the vortex[42].

In order to resolve the refined structure of the oscillations, we sweep the field at an extremely slow speed of 1 mT/min. $\bar{\eta}$ and $\Gamma$ in Fig. 2(b) remain the same values as those at high magnetic field when the superconductivity disappears. The oscillations of $\bar{\eta}$ and $\Gamma$ change to individual spikes with roughly the same height, each representing one single event occurs in the sample. In the zoom-in plot of $\Gamma$ in Fig. 2(c), we find that these events consist sharp decrease of $\Gamma$ followed by an exponential decay. We summarize the extracted event-intervals $\Delta B$ and their relaxation time $\tau$ in Fig. 2(e) and (f), respectively. $\Delta B$ and $\tau$ exhibit no clear dependence on the magnetic field. The histograms of $\Delta B$ and $\tau$ are fit using the normal and log-normal distributions in Fig. 2(g) and (h), respectively. The relaxation time for $\Gamma$ to decay back to the baseline after every jump is $\sim$ 30s, which is too slow to be caused by any microscopic process.

In type-II superconductors, the magnetization $M = -\frac{1}{4\pi}\left(H - \frac{n\Phi_0}{\mu_0}\right)$ is usually known as smooth functions of the external field $H$. Slowly increasing (decreasing) $H$ introduces current flowing along the edge of the sample through the Meissner effect, which generates a force to compress (or stretch) the vortices. As a result, the vortex density $n$ varies to minimize the Gibbs free energy. However, the strongly interacting vortices tend to form a crystal; see the white dots in Fig. 1(a). They are collectively pinned by the small but ubiquitous disorders and the crystal is effectively incompressible, i.e., $n$ remains constant. The super current flowing along the sample edge cancels out the effective magnetic field $H_{eff} = H - \frac{n\Phi_0}{\mu_0}$ experienced by the superconducting phase outside the vortex core. The whole system will exhibit perfect Meissner effect $M = -\frac{1}{4\pi} H_{eff}$ until $H_{eff}$ reaches $\pm H_{c1}$, when new vortices avalanche into the sample and destroy the vortex lattice; see the black dots in the third cartoon of Fig. 1(a). The randomly localized vortices then gradually re-crystalize into a new crystal with a different lattice constant. This process manifests as a sharp jump followed by an exponential decay in both $\Gamma$ and $\bar{\eta}$, see Fig. 2(b) and (c). The free energy reduces, and part of the released energy is transferred to SAW, causing a negative $\Gamma$. Similar phenomenon has been reported as ultrasonic amplification if the current carrying particles move faster than the acoustic wave[43].

This scenario is consistent with the fact that $\Delta B$, the field separation between two neighboring events, ranges from 0 to about $2H_{c1}$ (13 mT) with merely two exceptions, because $H_{eff}$ ranges between $\pm H_{c1}$ after every crystal reset. Its average value is about 6.6 mT, also in good agreement with the expected $H_{c1}$[34]; see Fig. 2(d). When we increase the field sweeping rate to 10 mT/min, the discrete neighboring events tend to merge together and show an approximate $B$-periodic oscillating behavior. The merging tends to occur in bundle of three events, which might be related to the triangular lattice structure of the vortex lattice. Such an evolution resembles a similar shot-noise of ordered (crystal) phase seen in two-dimensional electron system[44], where the electron solid generates random pulse-like current noise at the onset of depinning and quasi-periodic (narrowband) noise if the solid slides continuously[45, 46]. This scenario is also consistent with the observation that the oscillation fades out at large field, fast sweeping rate (Fig. 2(i)) or higher temperatures (Fig. S5). The large disturbance forbids the vortices from arranging into large crystals, but rather generates glass- or even liquid-like phases where new vortex can be generated smoothly[15–17]. Moreover, the non-equilibrium and chaotic vortex configurations no-longer have collective pinning so that the absorption of SAW also increases.

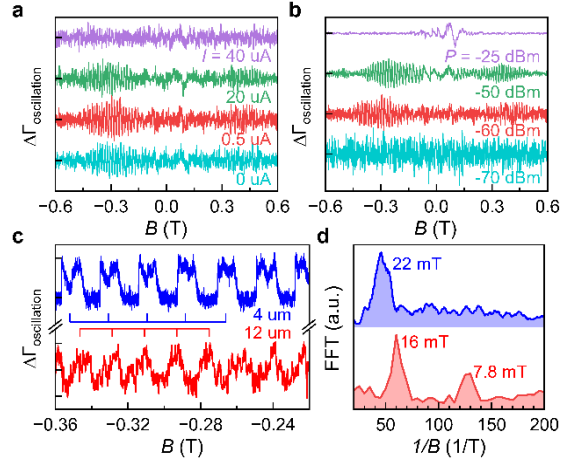

**Figure 3│Current, SAW power and IDT period dependence oscillations. a,** Extracted attenuation oscillation $\Delta\Gamma$ oscillation vs. $B$ with different current through the sample. Current will apply Lorenz force on vortex, which disturb the system. **b,** $\Delta\Gamma$ oscillation vs. $B$ with different SAW power. The oscillation behavior becomes clearer with increasing power until -25 dBm, which may be too large that disrupts the vortex system. **c,** Comparison of oscillations of devices with 4 µm period and 12 µm period. For a 12 µm SAW device, $\Delta\Gamma$ oscillation shows smaller period. **d,** Comparison of FFT results. Second harmonic frequency (corresponding to 7.8 mT period oscillation) can be seen in 12 µm device, which approximately complies commensurability with 4 µm device's oscillation period.

Following the above rationale, we would expect that a DC current through the sample will suppress the oscillation, similar to using a faster sweep rate. Fig. 3(a) shows the response of the oscillation when different DC current flows through the sample. The oscillation disappears when the bias current increases to 40 $\mu$A. The corresponding current density is $(J_s)_{ext} \sim 2.2 \times 10^3$ A/cm$^2$, several orders of magnitude lower than the superconducting critical current density. It inserts a Lorentz force $\boldsymbol{F}_L = \boldsymbol{J}_s \times \Phi_0 \boldsymbol{k} \times d \sim 4 \times 10^{-15}$ N to the vortex, small but sufficient to overcome the pinning potential. Fig. 3(b) shows the response of the oscillation using different SAW power with $dB/dt$ = 100 mT/min. The oscillation becomes stronger when the input SAW power increases from -70 to -50 dBm, and then gradually disappears if we further increase the power to -25 dBm. Meanwhile, a $I$ = 0.5$\mu$A current is injected into the sample for the conventional transport measurement and experimental results shows its superconductivity remains strong[35]. These current and power dependences can help to exclude many extrinsic factors such as the vibration of dilution refrigerator, the crosstalk of measurement cables and the deformation of sample holder under magnetic field. We measure another device with 12 $\mu$m SAW and the oscillation is weaker but still clear in Fig. 3(c). The 12 $\mu$m device exhibits clear second harmonic peak corresponding to ~ 7.8 mT, as shown in Fig. 3(d).

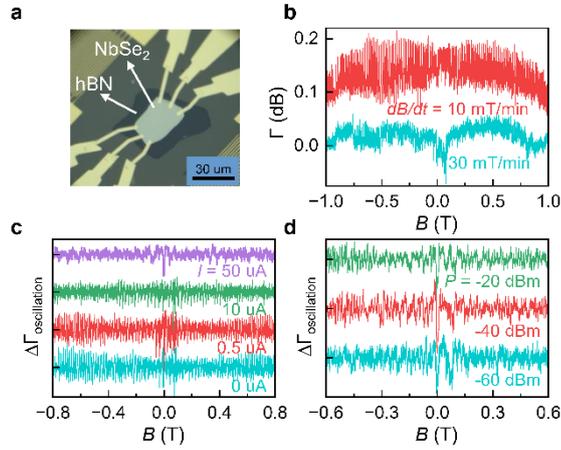

**Figure 4│Another sample shows the same phenomenon. a,** Photo of the NbSe2 device. 30 nm thick hBN is used to separate NbSe$_2$ from the LiNbO$_3$ substrate. **b,** ΔΓ oscillation vs. B with field-sweeping rate of 10 and 30 mT/min. The oscillation weakens for faster field sweep. **c,** Increasing current also weakens the oscillation. **d,** Changing SAW power at 100 mT/min sweeping rate. Intermediate SAW power will increase signal quality, while oscillation disappeared at large SAW power.

The oscillation can be ubiquitously observed in all clean type-II superconductors. In Fig. 4, we study a ~ 50-nm-thick NbSe$_2$, another superconductor sample whose $H_{c1}$(9 mT), $H_{c2}$(5 T) and $J_c$(1.6×10$^4$ A/cm$^2$) are very different from those of the PdBi$_2$[47–50]. The characteristics of the oscillation is remarkably similar to those of the PdBi$_2$ sample, despite their very different critical field/current, etc. A weak but clear oscillation is seen if the sweep rate is sufficiently slow. The oscillation becomes stronger at -40 dBm input power and disappears at -20 dBm. It persists up to 10 $\mu$A DC current and nearly completely disappears at 50 $\mu$A.

In summary, we use SAW to probe the superconducting vortex dynamics in weak pinning systems under magnetic field, and provide a new method for accurately measuring $H_{c1}$ of superconductors from the sound velocity plateau seen near zero magnetic field. We observe discrete, spike-form events in SAW velocity shift and attenuation which can only be observed when magnetic field varies. With increasing field sweeping rate, the events will merge into *B*-period oscillations which can be affected by other external conditions such as SAW excitation power and period, electric current. Our results provide new experimental evidences to understand the real-time vortex dynamics in superconductors and demonstrate SAW as an ideal technique to uncover rich physics of vortex systems.

**Methods:**

**Device Fabrication.**

We first prepared the 4 $\mu$m-period interdigital transducers (IDTs, Ti/Au with 8/50 nm thickness) on YZ-cut LiNbO$_3$ with standard electron-beam lithography (EBL) and metal evaporation. For PdBi$_2$ and NbSe$_2$ samples, we employed two methods to prepare the devices. (1) The $\beta$-PdBi$_2$ is obtained using Al$_2$O$_3$-assisted exfoliation[51]. We grow 60 nm of Al$_2$O$_3$ on the surface of freshly obtained PdBi$_2$ by thermal evaporation to assist in cleavage. The thickness of PdBi$_2$ flake was obtained as 90 nm by AFM measurement[35]. We picked up the PdBi$_2$-Al$_2$O$_3$ with polydimethylsiloxane (PDMS) stamp and landed it on the middle of two IDTs. The Al$_2$O$_3$ could separate PdBi$_2$ and LiNbO$_3$ to prevent any LiNbO$_3$'s interface effects. We deposited Cr/Au (5/100 nm) for electrical connections. (2) Due to being sensitive to air, NbSe$_2$ ($\sim$ 50 nm) is encapsulated with hBN (both $\sim$ 30 nm). We first transferred the bottom hBN on IDT substrate.

NbSe$_2$ was picked up and placed on the middle of the bottom hBN. Then we evaporated the electrode contacts (Ti/Au with 8/50 nm thickness) on top of NbSe$_2$. Finally the top hBN was transferred on top of the NbSe$_2$ (not shown in Fig. 4(a)). All the transferring processes are done in Argon-glove box with polycarbonate (PC)/polydimethylsiloxane (PDMS) stamp. During lithography and evaporation, NbSe$_2$ is placed in a custom-made vacuum box to avoid exposure to air.

**SAW and Transport Measurement.**

All the SAW measurement was measured on Oxford Triton 400 dilution refrigerator with the base temperature of 10 mK. Two 50 Ω resistances are connected in parallel to IDTs for broadband impedance matching. The frequency dependent transmission coefficient $S_{21}$ (shown in Fig. 1(b)) was measured with vector network analyzer (Ceyear, 3674H) to determine the SAW resonance frequency. We used a custom-made high-precision lock-in amplifier to measure the small SAW amplitude $A$ and phase $\phi$ changes[52]. The SAW velocity shift $\bar{\eta}$ and attenuation $\Gamma$ can be obtained with: $\bar{\eta} = \frac{\phi(B)-\phi_0}{2\pi}\frac{\lambda}{L}$ and $\Gamma = -20 log \frac{A(B)}{A_0}$, where $\phi(B)$, $A(B)$ and $\phi_0$, $A_0$ are the phase and amplitude of certain field and high field. $\lambda = 4~\mu$m is the SAW wavelength, $L = 358~\mu$m is the center-to-center distance between two IDTs. The transport resistance was simultaneously carried out with SAW measurement. We used Stanford Research Systems SR830 lock-in amplifier to perform low-frequency (13.333Hz) constant current measurement. Unless otherwise specified, the applied electric current is 0.5 $\mu$A, and the input excitation SAW power is -60 dBm (1 nW).

**Acknowledgments:** We acknowledge support by the National Key Research Program of China (Grant No. 2021YFA1401900, 2022YFA1403300 & 2020YFA0309100), the Innovation Program for Quantum Science and Technology (Grant No. 2021ZD0302602 & 2024ZD0300103), the National Natural Science Foundation of China (12074073) for sample fabrication and measurement. M.L.Tian acknowledges the support by the The Basic Research Program of the Chinese Academy of Sciences Based on Major Scientific Infrastructures (grant No. JZHKYPT-2021-08). We thank David Huse, Bo Yang and Fa Wang for valuable discussions.

**Author contributions:** Y.W. and L.G. conceived and designed the experiment. M.T. supplied the $PdBi_2$ material. J. Li and S. Jia supplied the $NbSe_2$ material. Y.W. and L.G. fabricated the samples. Y.W., L.G., and R.W. carried out the experiments, and performed the data analysis. Y.W., L.G. and Y.L. wrote the manuscript and Supplementary Materials with input from all authors. All authors contributed to the discussion and interpretation of the results. H.G., J.S. and Y.L. supervised the entire experiment and writing process.

**Additional information:** Additional information is available in the Supplementary Materials.

**Competing interests:** Authors declare that they have no competing interests.